\documentclass[aps,physrev,preprint,groupedaddress]{revtex4-2}

\usepackage{graphicx}

\begin{document}


\title{Generative AI as a lab partner: a case study}


\author{Sebastian Kilde-Westberg}
\email{sebastian.lofgren@physics.gu.se}
\author{Andreas Johansson}
\author{Jonas Enger}
\affiliation{Department of Physics, University of Gothenburg, SE 412 96 Gothenburg, Sweden}



\begin{abstract}
Generative AI tools, including the popular ChatGPT, have had a significant impact on discourses about future work and educational practices.
Previous research in science education has highlighted the potential of generative AI in various education-related areas, including generating valuable discussion material, solving physics problems, and acting as a tutor.
However, little research has been done regarding the role of generative AI tools in laboratory work, an essential part of science education, and physics education specifically.
Here we show various ways in which high school students use ChatGPT during a physics laboratory session and discuss the relevance of using generative AI tools to investigate acoustic levitation and the speed of sound in air.
Additionally, employing variation theory as a theoretical lens in the analysis, we identify how generative AI can be used to further develop students' problem-solving skills in the physics laboratory.
However, although our study shows that generative AI tools may handle some relevant questions and problems during laboratory work, the teacher still plays a crucial role in identifying students' needs and capabilities of understanding the potential and limitations of generative AI.
Finally, this study serves as an important point of discussion regarding the ways in which students need support and training to efficiently utilize generative AI to further their learning of physics.

\end{abstract}


\maketitle

\section{Introduction}
\label{sec:introduction}

The introduction of generative artificial intelligence (GenAI) tools, spearheaded by the public release of the chatbot ChatGPT \cite{ChatGPT}, has led to a broad debate on their impact on education and future workforce needs \cite{EUreport, Shen2024, Wells2024, worldeco}.
In physics education research (PER), recent studies involving GenAI tools have examined their ability to generate valuable (i.e., reliable or interesting, but sometimes incorrect) answers to physics problems \cite{Dahlkemper2023, Gregorcic2023, Kortemeyer2023, lopez2024}.
Although a need has been identified for physics instructors to incorporate artificial intelligence (AI) in the classroom, there is still little focus on how GenAI tools can be used in the physics laboratory, a crucial part of students' education when learning physics.

In this study, we investigate the potential future role of GenAI tools, like ChatGPT, in the context of high school physics laboratory.
We do this in part by analyzing students' awareness during the lab by drawing on the variation theory of teaching and learning as a theoretical framework.
Specifically, our work provides valuable insights into how students are utilizing the freely available version of ChatGPT, GPT-3.5 at the time of data collection, with varying prior knowledge of the phenomenon they are studying, as well as with little to no prior training on how to use GenAI tools in school.
Our aim is to contribute to the discussion on the ways in which GenAI tools may impact future laboratory work and implications for teachers, both in planning and executing such lessons.
We do this by answering the following research questions (RQs):

RQ1: \textit{What do students perceive as useful interactions with GenAI tools like ChatGPT during a lab session?}

RQ2: \textit{How do high school students actively engage with ChatGPT to solve problems during a lab investigating the speed of sound using acoustic levitation?}

RQ3: \textit{How can interacting with ChatGPT during the lab aid students in solving problems and with conceptual understanding?}

The structure of the paper is as follows. In Sec. \ref{sec:background}, we provide a broader context for the study by conducting a literature review to identify prior research that can help inform the use of GenAI in the physics laboratory.
Sec. \ref{sec:theory} describes the theoretical framework of variation theory, which is a theory of teaching and learning that was used in the current study.
By introducing a theory of teaching and learning, it provides a clear lens through which we can connect our results to practical implications for teaching and learning in the physics laboratory.
In the current study, we draw upon our theoretical framework specifically in analyzing the results as a way to answer RQ3.
Our lab design, as well as the study design, is presented in Sec. \ref{sec:method}, which also includes a discussion of the study's limitations and how the data were analyzed.
Our findings are presented in Sec. \ref{sec:findings}.
In Sec. \ref{sec:discussion}, the RQs are discussed in relation to our findings. There, we also discuss implications for teaching and learning in the science laboratory.
Finally, Sec. \ref{sec:conclusion} provides a summary of the paper, including an outlook that relates the findings of the current paper to future research needs in the field.

\section{Background}
\label{sec:background}
PER involving modern GenAI tools (post public release of ChatGPT) has had a broad focus, including comparing different models' performance on tasks (e.g. \cite{Polverini2024b, Fussell2025}), performance on physics problems (e.g. \cite{Kortemeyer2023, Yeadon2023, Yeadon2024, lopez2024, Polverini2024a, Pimbblet2025}), ways in which physics teachers can use these tools (e.g. \cite{Gregorcic2023, Kchemann2023, Domenichini2024, Gregorcic2024, Kortemeyer2024, Wan2024}), and the potential for interactions between students and GenAI (e.g. \cite{Dahlkemper2023, Ding2023, Liang2023, Kahaleh2025, Trout2025}).
A common theme that can be observed in PER is that the use of GenAI tools, whether for students, teachers, or researchers, must be carefully considered to achieve a beneficial outcome.
Regarding the potential use of GenAI tools for students, a primary benefit has been identified to be connected to increased motivation and engagement \cite{Adiguzel2023}.
However, a recent systematic review on the influence of ChatGPT on student engagement has found that the increased engagement can, in some instances, be linked to cheating and reduced critical thinking, resulting from over-reliance on answers provided by chatbots during problem-solving activities \cite{Lo2024}.
These findings reinforce the notion that the use of such tools in educational contexts requires careful consideration.

In science education, research about incorporating GenAI tools in laboratory work has primarily been done in the context of chemistry education, including comparing different models \cite{Hallal2023}, aiding students developing methods \cite{Ruff2024}, producing code to assist in data analysis \cite{Hare2024}, and in writing lab reports \cite{Humphry2023}.
Two studies, both conducted at the university level and utilizing ChatGPT with the freely available version, GPT-3.5, involved students who had little to no prior experience with GenAI tools \cite{Hare2024, Ruff2024}.
In using ChatGPT to help students produce code to analyze lab results, students found the GenAI tool to be useful and reported an increased level of understanding of the programming language they used \cite{Hare2024}.
The other study had students use ChatGPT to help them develop methods for lab proposals in analytical and inorganic chemistry \cite{Ruff2024}.
Results indicated that it was challenging to obtain useful information from ChatGPT about specific procedures, as it often provided some correct information alongside vague or incorrect information.
Additionally, the authors found that students with more theoretical and experimental knowledge were able to utilize ChatGPT more effectively than novice students.

A link between students' prior knowledge and their ability to evaluate the quality of information provided by GenAI tools has also been identified in PER, outside laboratory contexts.
In a study investigating students' ability to evaluate the scientific quality of answers to three physics problems, 102 first- and second-year undergraduate students were presented with four different answers to each problem, three incorrect and generated using ChatGPT, and one correct generated by the researchers \cite{Dahlkemper2023}.
Their findings identified that the students' ability to gauge the quality of the answer was closely linked to their prior knowledge of the underlying physics.

Finally, in addition to prior knowledge influencing students' ability to identify incorrect information provided to them by GenAI tools, their implicit trust in chatbots can also influence how they gauge answers they get when interacting with the tool. This was investigated in a study having students use ChatGPT as a tutor \cite{Ding2023}. They found that students preconceptions about the capabilities of GenAI tools influenced their trust in answers provided by ChatGPT. Additionally, they also identified that many of the participating students blindly trusted answers provided by ChatGPT, pointing towards students not having received much in terms of prior teaching about GenAI tools or how they are typically developed.

\section{Variation theory}
\label{sec:theory}
The current study is exploratory in nature, and the findings presented in this paper were identified using an inductive approach, similar to when applying grounded theory \cite{Glaser2017} as a research approach, or conducting a thematic analysis \cite{Braun2006}.
In this section, we introduce the variation theory of teaching and learning as a theoretical framework to be used in discussing our findings, allowing for further analysis related to students' awareness, as well as discussing general principles on teaching and learning that the theory identifies.
As such, variation theory provides a lens through which we can discuss the implications of our findings for teaching and learning in the science laboratory.

Variation theory is a theory of teaching and learning asserting that for learning to take place, the learner must notice what is to be learned, and that noticing requires variation \cite{Marton2004, holmqvist2014, Marton2015, Kullberg2024}.
It argues that it is easier to notice what differs between two contexts than what is similar, and learning is to be understood as `learning to see' \cite{Marton2015}.
Here, we can identify that learning is closely connected to a learner, meaning that in variation theory, one must acknowledge that the process of learning about a specific concept or phenomenon within a particular subject can not be viewed as an isolated event, void from individual differences, including previous experiences of what is to be learned.
Rather, a teacher must account for the students when designing teaching sequences, because students' prior knowledge and experiences of the world affect how they discern the current content.

\subsection{Object of learning, structure of awareness, and critical aspects}

What is to be learned is referred to as an object of learning.
It is related to a specific group of students and the surrounding educational context, including curriculum-imposed goals \cite{Marton2015, Kullberg2024}.
An object of learning is constituted by a direct and an indirect object of learning.
The content (e.g., standing waves) to be learned refers to the direct, whereas a particular skill or capability to be developed (e.g., predicting how particles can be trapped using acoustic levitation) refers to the indirect object of learning \cite{Marton2015}.
Together, they make up the whole object of learning, e.g., using the concept of standing waves to explain and predict how to trap particles using acoustic levitation, which is then identified as what is to be learned.

For a particular object of learning, one must also consider how a particular group of students may initially perceive or discern it.
How a student perceives an object of learning is related to their current structure of awareness, an idea stemming from the phenomenographic tradition \cite{Marton1997}.
In the phenomenographic tradition, it assumes that a learner's structure of awareness is their experiences of the world.
The structure of awareness continuously changes, as humans can only be consciously aware of a limited set of the world simultaneously.
When considering an object of learning, we are thus never capable of being actively aware of all parts, or aspects, of it at the same time.
As a student, what is to be learned can, from this point of view, be understood as becoming aware of parts of the object of learning in more powerful ways.
These parts, that the student has yet to discern, are known as critical aspects \cite{Marton1997, Pang2016}, and the structure of awareness dictates how students become aware of these aspects \cite{Hasan2024}.
Finally, the object of learning can be divided into an intended (the teacher's aim prior to the current lesson), an enacted (what aspects are made possible to discern during the lesson), and a lived object of learning (what the students actually learned).

\subsection{Putting it into practice}

When planning teaching and learning situations, variation theory emphasizes that focus should be on how variation can be used in handling the content that is taught, not variation in terms of varying teaching methods, such as types of classroom activities \cite{Kullberg2024}.
By varying the handling of the content, one can highlight the critical aspects of the object of learning in ways that make them more clearly visible to the students.
However, if the teacher is not cognizant of their students' current possible structure of awareness, introduced variations may shift their focus towards unproductive, misleading aspects of the phenomenon under study.
Specifically, during a lesson, students are ideally better at discerning critical aspects of the phenomenon under study towards the end of the lesson.
This means that the students have become more proficient in identifying what the important (critical) aspects are.
The students' capacity to better connect the critical aspects to the object of learning means that they have developed a more refined awareness of the phenomenon \cite{Hasan2024}.
If students become aware of the intended object of learning during the lesson, it also means that the enacted object of learning is in line with the intended.
Such an alignment relies on the students discerning the critical aspects, but also that the students' structure of awareness is directed towards the object of learning.

Studies employing variation theory commonly focus on analyzing qualitative data, and the same holds for studies using this theory in PER (e.g., \cite{Ingerman2009, Euler2020, KildeLfgren2023}).
The three studies mentioned here to exemplify the use of variation theory in PER employ similar setups in that they follow a study design where a lesson has been designed.
However, \cite{Ingerman2009, Euler2020} both have one set intervention design, whereas \cite{KildeLfgren2023} iterates the design throughout the study, following a design-based research approach to study students' understanding and improve the lesson design.
Results from all three studies provide fruitful insights into what learning is made possible with each specific lesson design, and how teachers' intervention can aid in further directing students' attention towards critical aspects.
On a more general note, they also highlight how variation theory can serve as a powerful tool in designing and planning good teaching and learning sequences in physics.

In the context of the current study, variation theory allows us to identify and discuss the implications of the findings presented in this paper for designing fruitful teaching and learning sequences in the age of AI in education.
This is done in part by identifying instances during the lab where students' awareness shifts, indicating that their conception of the object of learning, or some critical aspect(s), has changed.

\section{Method}
\label{sec:method}

The current study is designed to identify different ways students might naturally find it useful to make use of GenAI tools readily available to them in a laboratory situation.
To that end, we opted to let students have access to one of the most common ways they might have come in contact with AI, the current free version of ChatGPT.
Further, a lab was designed with the research questions in mind.
As such, we intentionally focused on constructing a lab session so that students had little to no direct knowledge about the phenomenon under study and would struggle to complete the lab without consulting ChatGPT.
In this section, we provide an introduction to the lab that was designed, as well as the context of the study in terms of participants, ChatGPT, data collection, and the analysis process. Finally, limitations connected to the chosen approach are also discussed.

\subsection{Experimental setup and lab design}

\begin{figure}[b]
\includegraphics[scale=1]{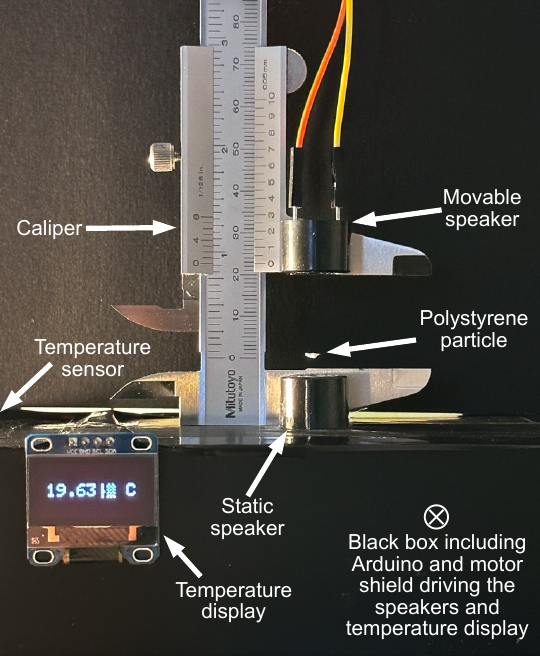}
\caption{\label{fig:exp_setup} The experimental setup, also known as LeviLab. The external low-voltage DC power supply powering LeviLab is not included in the figure.}
\end{figure}

In acoustic levitation, particles typically in the millimeter range are trapped mid-air due to the creation of an acoustic field of standing waves.
Millimeter-sized particles consisting of materials denser than air are attracted to the sound intensity nodes \cite{Johansson2024a}.
The experimental setup (LeviLab, \cite{Johansson202b}) comprises two ultrasonic speakers mounted vertically and opposite to one another on a caliper (Fig. \ref{fig:exp_setup}). 
As the speakers, which are connected in phase, are moved axially, the standing waves between them shift.
If a millimeter-sized object is placed resting on the bottom speaker grill and the top speaker is slowly moved closer, the particle will be attracted to the closest node, which can either be above or below the speaker grill.
This results in that the particle will, at certain distances when a node above the speaker grill is sufficiently close, seem to jump up and begin to hover mid-air.
The configuration allows for the variable and precise measurement of the distance between the loudspeakers.
In addition, LeviLab outputs a constant frequency of $40 \text{ kHz}$, and is equipped with a thermometer and a digital display, which allows a measurement of the ambient temperature while performing experiments using the setup.

Additionally, the students were encouraged to also bring their physics textbook, formula book, and any calculator. The lab manual guided the students in obtaining data to analyze, from which they could calculate the wavelength and the speed of sound.
The method for data collection was precisely described. However, the instructions related to the analysis had been intentionally designed to be somewhat vague, asking students to perform a linear regression and to calculate the speed of sound using a formula that includes wavelength and frequency without explicating the exact form of the equation.
Additionally, students were also tasked to compare the speed of sound with another model where the speed of sound depends on temperature. The intentional vagueness made it so that all or most groups would have to consult ChatGPT to identify what equation to use, since it was beyond their expected prior knowledge, regardless of educational level.
Lastly, to increase the need to consult ChatGPT, discussion questions were added, asking the students to construct a figure that describes how sound waves can be used to counteract gravity, potential sources of errors, the relevance of measuring temperature, and regarding applications of acoustic levitation.
 Students were considered to have successfully finished the lab if they finished the analysis, even if not all discussion questions were answered.

\subsection{Participants}

The current study was conducted with Swedish second- and third-year high school students taking one of the three physics courses offered in the Swedish high school curriculum \cite{Skolverket2011}.
Having students taking different level physics courses was made possible since the phenomenon of acoustic levitation is not present in the Swedish high school or compulsory school physics curriculum.
However, in compulsory school, students learn about the physics of sound, specifically related to how it originates, propagates, can be reflected, as well as the particle model of matter including, e.g., pressure and density \cite{Skolverket2024}.
In the high school curriculum, however, sound is not covered in the course Physics 1, but again brought up in Physics 2 as part of learning about mechanical waves, and in Physics 3 in more advanced study of mechanical waves \cite{Skolverket2011}.
Thus, we expected the participants to have different, but at least some, knowledge about sound regardless of what Physics course they are currently taking.

Participating students taking Physics 1 or 2 were sought at one high school, and students taking Physics 3 were sought at a university that offered the non-mandatory Physics 3 course to third-year high school students.
The sampling of students was done by convenience sampling, where all willing students got to participate in the study.
In total, seven groups participated: four with second-year students taking Physics 1, one with third-year students taking Physics 2, and two with third-year students taking Physics 3.
All groups consisted of three students, except Group 7 (Table \ref{tab:participants}).
\begin{table}[t]
\caption{\label{tab:participants}%
A summary of each lab group, including pseudonyms for each of the participants ($n=19$), educational year, what Physics course they are taking, and their effective lab time rounded up to the nearest minute.
}
\begin{ruledtabular}
\begin{tabular}{rlrrr}
\textrm{Lab group}&
\textrm{Students\footnote{Pseudonyms that preserve anonymity regarding name and gender by choosing names from the list of the most common baby names in Sweden in 2005 \cite{scb}, half boys and half girls. Names were organized and chosen in alphabetical order, with one exception in E, where the first boy's name, Ebbe, was skipped due to the similarity to Ebba.}}&
\textrm{Educational year}&
\textrm{Physics course}&
\textrm{Lab time (minutes)}\\
\colrule
1 & \textrm{Adele, Adam, Agnes} & 2 & 1 & 43\\
2 & \textrm{Benjamin, Beatrice, Bill} & 2 & 1 & 44\\
3 & \textrm{Caroline, Carl, Cassandra} & 2 & 1 & 56\\
4 & \textrm{Daniel, Daniela, Dante} & 2 & 1 & 55\\
5 & \textrm{Ebba, Eddie, Edith} & 3 & 2 & 46\\
6 & \textrm{Fabian, Fanny, Felix} & 3 & 3 & 47\\
7 & \textrm{Gabriella} & 3 & 3 & 56\\
\end{tabular}
\end{ruledtabular}
\end{table}
Furthermore, since each lab group participated in the study parallel to their normal lesson, there was some variance in the time each group had for the lab and interviews.
On average, the pre and post-interviews took five minutes each and between five to ten minutes was required for transport to the classroom and informing the participating students about the study.
In total, each group was given between 65 and 70 minutes to participate in the study.
The effective lab time, which was either until they declared they were finished or the researcher present in the room had to stop them, for each group is presented in Table \ref{tab:participants}.

Ethical considerations for this study have been made following local rules and guidelines set by the Swedish Research Council \cite{SwedishResearchCouncil2024}. All participating students were old enough to understand the purpose of the study and the data collection method. Participants all gave written consent after being informed of the purpose of the study, what data was being collected, how it was to be handled to preserve their anonymity, and that participating was voluntary.

\subsection{ChatGPT}

Since the behavior of ChatGPT is changing over time, even for the same model \cite{Chen2024}, here we provide context regarding what model was used during the study, as well as when it was used.
The data collection was done between March 5 and March 28, 2024.
During each lab session, students had access to the currently freely accessible version of ChatGPT, GPT-3.5.
Further, between each lab session, the entire chat history of the account was downloaded and then cleared from the account, meaning each lab group faced a seemingly empty account with no access to previous chats or any initial prompting done by the researchers.
The account students used ChatGPT from was one created by the researchers.

We opted for the free version of ChatGPT after carefully considering its ability to assist with the analysis during the lab and the types of answers it could provide for the optional discussion questions towards the end of the lab.
In our testing of GPT-3.5 in February 2024, we identified that the current version of ChatGPT could be a sufficient help in that it was able to provide answers, albeit of varying quality depending on how the prompts were constructed, to important questions related to the analysis students were tasked to conduct during the lab.
Another argument for using the free version of ChatGPT during the study was that it would more closely perform on the level students who had prior experience with ChatGPT or similar GenAI tools thought it would.
Thus, their use of ChatGPT during the study would be similar to how they would use it in a more naturalistic setting.

\subsection{Data collection}

To get relevant information about how the students worked during the lab, each group was filmed during the entire lab session, including the interview before and after the lab.
Additionally, the entire chat history with ChatGPT was collected, as well as any written notes the groups created during the lab session.

During the interview before the lab session, students were asked if they had used ChatGPT or similar services before for school-related tasks, with a follow up on use in physics specifically.
If they had such experiences, then follow-up questions asked for examples and what version of ChatGPT or other tool they had used.
Then, to get a sense of the students prior knowledge and how they approach solving problems during labs, we also asked them what they think acoustic levitation is and how they would handle a situation where they run into a problem, practical or theoretical, during a normal lab session.

\begin{figure}[b]
\includegraphics[scale=1]{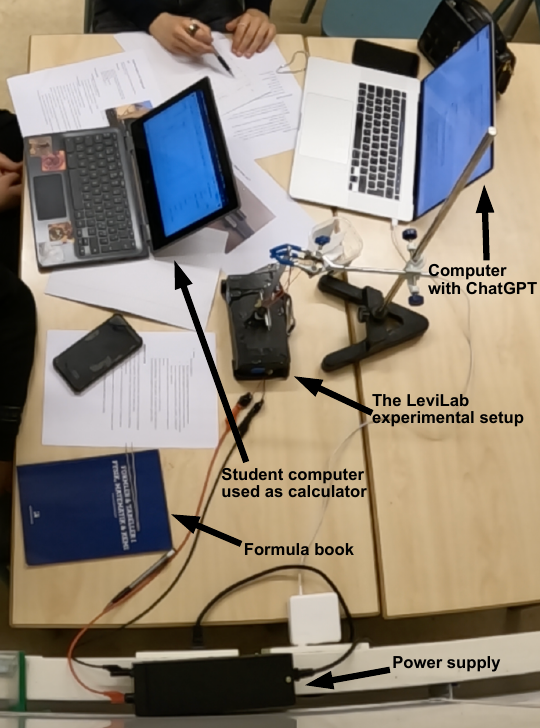}
\caption{\label{fig:workspace} What the students' laboratory workspace typically looked like, and what materials they chose to use.}
\end{figure}

During the lab, the students were told that the researcher present in the room during the lab session would not interfere or answer any questions unless something unexpected happened with the equipment.
In addition to the material needed to conduct the experiment during the lab session and what material students decided to bring to the session, a computer was also supplied by the researcher with ChatGPT.
A typical view of their workspace space can be seen in Fig \ref{fig:workspace}.
The researcher would also interfere if students' handling of the equipment would have led to breakage or posed any danger to the students.
Interference from the researcher was needed in groups 1, 2, 5, and 6 for about one minute within the first five minutes of the lab session.
Groups 1, 2, and 5 had trouble understanding how to use the caliper to vary the distance between the speakers, and it was deemed necessary for the researcher to explain this as they were at risk of breaking the glue holding the caliper and speakers in place.
Group 6 experienced glitches with the cables between the setup and the power supply, and the researcher had to step in to provide new cables and perform a quick check to see if any of the electronics had been damaged.
Finally, since Group 7 consisted of only one student, the researcher informed the student that they might ask clarifying questions about what they were doing during the lab session if it was deemed difficult to interpret what the student was working on only from the recorded video.
The researcher asked questions regarding what the student did after 16, 40, and 50 minutes, respectively.

After the lab, another group interview was conducted in which the students were asked to first think aloud regarding how they thought about using or not using ChatGPT during the lab.
That question was then followed up by asking if they discovered any problems in the answers they got from ChatGPT during the lab.
Then, the students were asked to reason about if they think tools like ChatGPT can be used to get help with explanations in physics, as well as if they can use such tools to get help with various problems during a lab session.
Finally, the groups were asked how they think AI tools could be used during lab sessions in physics.

\subsection{Analysis and limitations}

The research questions for this study are of an exploratory nature, with the aim of getting information about students' views of the potential role of GenAI tools in the physics laboratory, as well as furthering our understanding of the potential use cases of such tools in educational lab sessions.
By collecting qualitative data and identifying that the three parts of the study, the pre and post-interviews and the lab session, provide a narrative account of students' experiences as a group, the analysis focused on extracting these by drawing upon an inductive narrative approach for the analysis.

Following the typical structure of working with qualitative data, including conducting a narrative analysis \cite{crossley2007}, the analysis includes multiple steps.
First, to get familiar with the data, all of the video recordings were observed first as a whole, followed by a second time where the data was carefully transcribed.
The transcription process of the video recordings was aided by first having a local version of Whisper
\footnote{Using the app Ragnar, an open-source project available on Github: https://github.com/mickekring/ragnar.git}
transcribe the data.
Then, the first author certified that the transcription provided by Whisper was correct, adding or changing wording as needed, anonymizing the transcript, and formatting the transcription so it was possible to identify the different speakers.
After the transcription was complete, the first step of getting familiar with the data also included identifying what the groups did and when during the lab session.
This was done by constructing Gantt charts (Fig \ref{fig:gantt}).

The second step of the analytical process included inductively identifying relevant themes and passages that were in line with the general research aim of investigating the use of GenAI tools.
Here, it was relevant to make use of the Gantt charts, transcripts, and each group's chat history with ChatGPT to conclude first what they had done when, how they discussed interactions, or lack thereof, of ChatGPT, as well as connect why they answered the way they did during the post-interview in light of their reasoning during the pre-interview and interactions with ChatGPT during the lab.
Following the second step, parts three and four of the analysis included identifying the overall narrative tone and relevant themes and data to include in the findings sections.
Here, tentative findings were discussed both among all authors of the papers as well as presented and discussed with colleagues external to the current study.
Then, the main author proposed an initial structure of the findings and the main conclusions to be presented, which were again discussed among all authors to certify that the identified conclusions are supported by the data.
Practically, this was achieved by proposing themes and supporting transcripts, which were then discussed in terms of their reliability and potential problematic subjective interpretations.
After reaching a consensus among the authors on the main emerging themes identified in the data, the final analytical step involves reporting and discussing the findings, as presented in Sec. \ref{sec:findings} through Sec. \ref{sec:conclusion} of this paper.

In discussing the findings, the theoretical framework of variation theory was used in further analysis, related to the students' awareness.
Specifically, we relied on the structure of awareness as a way of understanding, from the students' point of view, the enacted object of learning.
The setup of the study, where no teacher was present to enact the intended object of learning, made it relevant to conceptualize the enacted object of learning.
In our analysis, we treated the enacted object of learning as something dynamic, akin to the dynamical nature of the structure of awareness.
As such, we could identify differences in what the enacted object of learning was between the groups, which aided our understanding of why the groups had seemingly different foci during the lab.
Additionally, variation theory was used as a lens in the discussion for identifying implications for teaching and learning.

Collecting and reporting qualitative data, as is done in the current study, is an important step on the path of creating a collective understanding of the phenomenon under study \cite{Cortazzi_1994}.
As with all qualitative research, however, it is important to recognize how the researcher influences the analysis, which brings to light the issue of validity \cite{crossley2007}.
Another issue that relates to the overall validity of the current study is the lack of representability and thus generalizability.
It is, therefore, important to recognize that the current study only included students enrolled in the Swedish school system.
As such, any conclusions drawn from this study have to be viewed in the light of the current overall cultural context of that system.
For example, it is important to recognize that soon after the public release of ChatGPT, the Swedish National Agency for Education urged schools to be cautious in letting students use such tools and to limit home assignments in general, as it would be nigh impossible to verify if the student or some GenAI tool had done the work handed in.
The level of knowledge the involved researchers had about the current local educational landscape, as well as the societal debate, was therefore deemed relevant to be cognizant of when conducting the current study.

Finally, it is important to again stress what conclusions and insights can be gained from the current study. It is not to provide findings to produce an objective truth or provide absolute certainty, but rather to identify and argue for justifiable conclusions that produce likely interpretations that have some utility for the intended reader \cite{crossley2007}.

\section{Findings}
\label{sec:findings}
Here, we present findings from the group interviews before and after the lab and what the students did during the lab, providing relevant data for answering the research questions.
The current study is a case study involving a small number of participating students ($n=19$), and as such, the findings presented here should not be seen as directly generalizable.
Further, in this section, we include prompts by the groups but give a more descriptive take on what ChatGPT answers.
We opted for this approach since all groups, except Group 6, interacted with ChatGPT in Swedish, and the act of translating the answers could have a profound impact on how the answers are interpreted by a reader.
The entire chat history for each group is provided as Supplemental Material. As such, the interested reader may translate the chat history to get a more complete context of each group's chat history.

\subsection{Group interviews}

In analyzing the interviews before and after the lab, we primarily sought to answer RQ1 and provide a potential additional perspective on some parts of the lab, namely the students' self-reflection of what they had done and why.
This additional information could be helpful to discuss RQ2 and RQ3.

Regarding what interactions with ChatGPT students perceive as useful in the context of a lab session, participants were primarily negative about the general usefulness of ChatGPT in physics and mathematics before the lab.
All groups did mention that AI chatbots like ChatGPT were not good at math but could help provide comprehensible explanations of ``words" (concepts) they had forgotten or thought the book or teacher did a poor job explaining.
However, groups 4, 6, and 7 all stressed the importance of being cautious regarding the facts provided by ChatGPT and consistently trying to double-check with other reputable sources.
A few students (Benjamin and Beatrice in Group 2, Caroline in Group 3, and Daniel in Group 4) had no prior experience using ChatGPT in physics prior to the current study.
Of those students, Beatrice claimed they had not used AI chatbots for any school work previously, whereas Benjamin, Caroline, and Daniel mentioned having used it for help in other subjects that were not science or math.

After the lab, all groups except Group 5 were still skeptical of its ability to perform calculations or do any meaningful analysis.
Apart from Group 5, all students mentioned that it might not be a good idea to rely on ChatGPT for calculations.
Groups 4 and 6 said ChatGPT was helpful during the analysis since it could provide relevant equations and that they could trust them by either studying the equations and see if they looked correct or by performing calculations and verifying if the answer was in line with their hypothesis.
However, Group 6 specifically mentioned that this requires that you know something about the underlying physics.
Further, Group 3 said they got an equation from ChatGPT to use during the analysis, but that they could not know if it was correct since ChatGPT did not provide any source.
Within group 5, the reasoning was that ChatGPT was useful when one has to do 
\textit{``simple analysis that only takes time to write down and calculate. For that, ChatGPT is useful and good to use ... it gives a good estimate based on previous research and is therefore a good base answer."} (Ebba, group 5).

The students who had no previous experience with using AI chatbots in physics, Benjamin, Beatrice, Caroline, and Daniel, stated after the lab that they would use such tools in the future when they are looking to get short and simple explanations of things they struggle to understand.
Furthermore, most groups stated that their answers from ChatGPT during the lab seemed reliable and felt no need to fact-check theoretical answers.
Groups 4, 6, and 7 had a similar stance on trustworthiness before and after the lab regarding theoretical information. However, all mentioned that they thought it was reliable in providing physics formulas because they did not see any reason for fake information about those in the training data.
Fabian in group 6 explicated this as \textit{``We got two different formulas for the speed of sound. The first gave like 2000 meters per second and the other 344 comma something, so it felt more reasonable. Often there is some truth in all it says because it always comes from some context, but it might be a context we don't know about. For example the first formulae I still think is right because it makes no sense to keep fake physics in the training} [data]\textit{, but it might be that it gave wrong units or something."}

When asked about how they thought AI tools like ChatGPT could be utilized in physics labs in the future, all groups, except group 5, said it could be a good substitute for the teacher in answering conceptual questions. Group 5 thought it could also be good to help with calculations or get answers to some practical questions if they did not have enough time to finish the lab during class properly. Gabriella (group 7) explicitly mentioned the problem of using AI tools to cheat but stated that \textit{``ChatGPT can help you formulate explanations you are not fully done with. It is less obvious cheating in labs because it can help you with explanations or clarifications, but never do the work for you."}
Additionally, groups 4, 6, and 7 were all of the mindset that AI chatbots could be a valuable tool during labs in the future because they can help students \textit{``get answers to small stuff like helping you remember some word or technique and help the teacher by outsourcing simple or `stupid' questions. But it's also important that students get better knowledge about how AI tools work and how to ask good questions."} (Fanny, group 6).
In relation to ``stupid" questions, no group clearly defined what they meant by similar wording, but most students mentioned that it could be useful to ask ChatGPT about the meaning of non-physics words that appeared on lab instructions that they had forgotten or to ask about concepts they should know about, but had forgotten since they had not heard the word for some time.

\subsection{Interactions with ChatGPT during the lab}
\begin{figure*}
\includegraphics[scale=0.28]{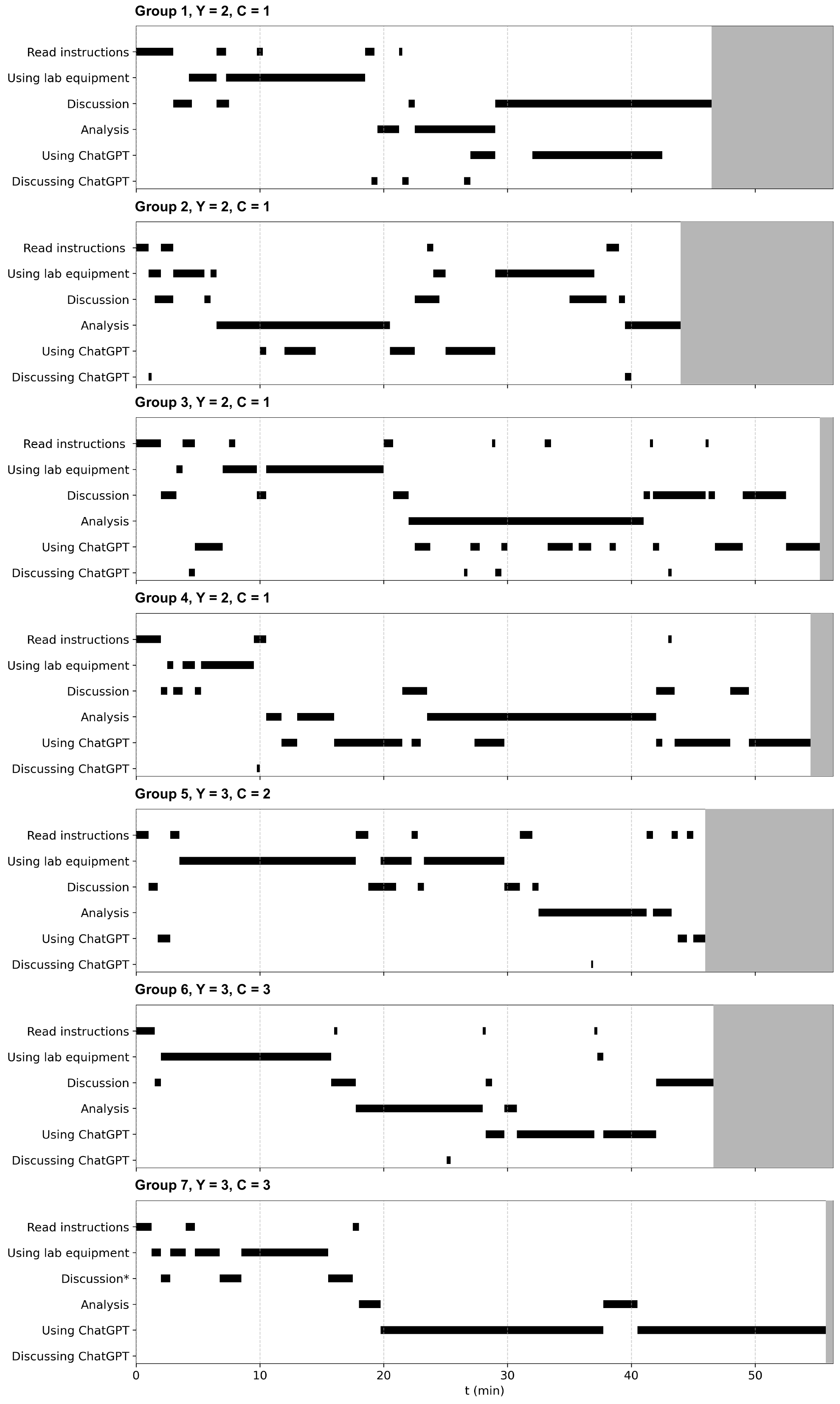}
\caption{\label{fig:gantt} Visualizing the workflow for each group, including information about their year (Y) and course (C), during the lab using Gantt charts using the categories defined in Table \ref{tab:categories}.
Since group 7 had only one student, the criteria for Discussion* was extended here also to include times when the student's actions did not fit in the other categories.
An example is when the student conducted more general note-taking during the lab. Gray bars are added to indicate each group's end time, as specified in Table \ref{tab:participants}.}
\end{figure*}

In this paper, we define one interaction with ChatGPT as consisting of one written prompt by the user and the reply from ChatGPT.
During the lab, the number of interactions the groups had with ChatGPT varied from two to 17.
Fig. \ref{fig:gantt} illustrates the manner in which each laboratory group engaged with the assigned tasks, categorized according to the following parameters: Read instructions, Using lab equipment, Discussion, Analysis, Using ChatGPT, and Discussing ChatGPT.
The meaning of the categories is elucidated in Table \ref{tab:categories}.
For each lab group, Fig. \ref{fig:gantt} indicates how they worked during the lab using the categories Read instructions, Using lab equipment, Discussion, Analysis, Using chatGPT, and Discussing ChatGPT (see Table \ref{tab:categories}).
\begin{table*}[b]
\caption{\label{tab:categories}%
Categories and corresponding criteria used to identify the overall workflow for each group during the lab.
}
\begin{ruledtabular}
\begin{tabular}{lp{12cm}}
\textrm{Category}&
\textrm{Criteria}\\
\colrule
\textrm{Read instructions} & \textrm{If students seem to be actively engaged in reading the lab manual}\\
\textrm{Using lab equipment} & \textrm{If students seem to be actively engaged in reading the lab manual}\\
\textrm{Discussion} & \textrm{If students talk among each other about the lab such as what to do, how to interpret results, sorting out questions, or talk about something that does not fit in the other categories}\\
\textrm{Analysis} & \textrm{If students are performing calculations or are otherwise involved in tasks related to data analysis, such as generating relevant graphs, as well as discussing or looking up formulas to use}\\
\textrm{Using ChatGPT} & \textrm{If students write to ChatGPT, read responses, or directly discuss responses or the interaction as a whole}\\
\textrm{Discussing ChatGPT} & \textrm{If students discuss something where ChatGPT is mentioned (or if any abbreviations such as ``Chat," ``the AI," or similar wording are used), without interacting with it}\\
\end{tabular}
\end{ruledtabular}
\end{table*}

Two groups, 2 and 5, tried to get some guidance from ChatGPT regarding data collection and equipment handling.
Group 5 tried to get practical guidance on how to do something with the lab equipment using the prompt \textit{``How do you adjust sound frequency on a Levilab"}.
This was asked after having read the instructions once and interpreting the instruction of varying the distance between the speakers to be synonymous with adjusting the sound frequency.
The response was seemingly irrelevant and mentioned a product called Levilab. However, it differed from the equipment they had access to, and the students re-read the instructions without interacting with ChatGPT until the end of the lab session.
Group 2, who struggled to identify what data to collect during the lab and, as such, had to collect data twice, did consult ChatGPT on \textit{``how DO YOU PERFORM LABORATIon of acoustic wavelength and sound speed"} about 20 minutes into the session.
For them, even though the reply from ChatGPT did not describe how to conduct measurements with the available equipment, the answer led them to understand that the concepts of sound waves, speed, and frequency were relevant to the experiment.

Groups 1, 3, 4, 6, and 7 had no interactions with ChatGPT connected to the data collection.
During the data collection process, Carl, in group 3, who was not involved in handling the equipment or taking notes, was tasked to ask ChatGPT to guide the group on what acoustic levitation does.
He did this first by asking \textit{``What is acoustic levitation?"}, with a follow-up to get more relevant information that was \textit{``Explain to a high school student"}.
Overall, the group was content with the second answer. There, ChatGPT used a metaphorical invisible basket and explained that it was created by special patterns of sound waves that made several pressure points that could be used to hold an object in place.
ChatGPT additionally made an analogy to trying to levitate balls in the air by blowing on them from below.
By accepting this explanation, the experimental setup the students worked with does not make much sense as it uses two opposing speakers to make a particle levitate, and the fact that the particle only levitates for some distances between the speakers
\footnote{Rather, acoustic levitation utilizes the phenomenon of standing waves to trap particles in sound intensity nodes. As a visual example, see Fig. 3 in \cite{Johansson2024a}}.
The answer ChatGPT gave to the first question Carl posed only mentioned that acoustic levitation uses sound waves to lift and manipulate objects and that sound waves can create an acoustic field that generates pressure points in a volume of air or other medium.
Thus, it was similarly vague regarding what happens, but omitted the use of analogies.

Group 6 also tried to get an introduction from ChatGPT on acoustic levitation.
However, the group was more familiar with mechanical waves, including relevant concepts such as standing waves, and utilized prior knowledge in their interactions with ChatGPT.
Additionally, they chose to interact with ChatGPT using English, as they claimed it had a lot more English than Swedish in its training set.
From the prompt \textit{``could you give us a introduction to acoustic levitation?"} they got a reply that involved relevant concepts such as standing waves, nodes and anti-nodes, and acoustic radiation pressure.
However, the claim that objects can be levitated in both nodes and anti-nodes did not agree with the group's hypothesized explanation of why the particle did not levitate at all speaker separation distances.
As such, they continued to probe if ChatGPT gave what they could gauge as a legitimate explanation of this claim by asking \textit{``Where in the standing wave can you balance an object and why?"}.
The new response from ChatGPT made them more confident in their hypothesis that the particle could only levitate in the displacement nodes since it once again mentioned levitation to be possible in both nodes and anti-nodes but also stated that levitation in the anti-nodes would be unstable but would still be attracted there due to a \textit{``stronger radiation pressure"}.
Group 6 discussed that this explanation was confusing and possibly based on flawed reasoning, although they were not sure because they felt they lacked the proper knowledge about the physics involved.

In general, groups 1, 2, 3, 5, and 7 were identified as struggling during the lab, meaning that they had trouble understanding what data to collect, got stuck during the analysis, or both.
For groups 1, 2, 3, and 5, interactions with ChatGPT focused on asking simple questions and they did little to no follow-up.
Group 7 struggled mainly due to a lack of knowledge on how to use the online calculator Desmos.
They opted for using Desmos since they did not bring a calculator to the lab session, and thus, much of the interaction with ChatGPT focused on trying to get help with understanding how to use Desmos to perform the necessary analysis.

Groups 4 and 6 both did well during the lab and showed a deeper understanding of how to interact with ChatGPT during the lab.
Questions they posed also included some poorly formulated ones such as \textit{``which sl unit is wavelength written with?"}.
However, a meaningful difference was the amount of follow-up and additional probing the interactions included.
Group 6 notably chose to double-check the answers given by ChatGPT by opening up additional chats and asking similar questions again but without having the context of the entire chat history.
Both groups also tried to more deeply understand how acoustic levitation works by posing different questions to ChatGPT, then discussed the answers among them and posed a new question to see if their understanding aligned with the new answers provided by ChatGPT.

\subsection{On interacting with ChatGPT and students awareness}

Drawing upon our theoretical framework, we can identify instances where the students' interactions with ChatGPT during the lab impacted their structure of awareness, which in turn has an effect on what the object of learning is (and thus what aspects are relevant to consider).
As an initial step, it must be acknowledged what students perceived to be the primary focus at the beginning of the lab session.
An aim for the lab session was provided to students by including a short sentence in the lab manual (see Appendix A) stating that they should explore acoustic levitation and determine the speed of sound in air.
During the data collection phase of the lab, both groups 3 and 6 verbally reiterated the written aim.
This led to two different interactions with the goal of better understanding acoustic levitation, where Group 6 discussed among themselves and reasoned about the phenomenon of acoustic levitation based on their prior physics knowledge.
Group 3 instead approached the aim of exploring acoustic levitation by having Carl interact with ChatGPT, as described above.
For the other groups, no discussions touching on the aim of exploring acoustic levitation were present until either during the data analysis phase of the lab, or after having finished the analysis.
Among these, Group 4 stood out by having conceptual discussions on why the particle could levitate during the data analysis stage.
Overall, most students perceived the aim of the lab session to be related to determining the speed of sound, and thus, the enacted object of learning was identified to be more closely connected to the capability of determining the speed of sound using the available equipment.
In light of the identified enacted object of learning, it was relevant for the groups to spend most of their time on data collection and analysis, as illustrated by Fig. \ref{fig:gantt}.

From the previous subsection presenting how groups interacted with ChatGPT, it became evident that the interactions were akin to having quick access to the teacher, a teaching assistant, or web searching capabilities.
In these interactions, we can identify passages where students' awareness of what could be important to discern during the lab was seemingly affected.
Here we present three examples, from groups 2, 4, and 5, highlighting how the interactions affected the students' structure of awareness during the episodes.

The first episode is of Group 2 and an initial interaction with ChatGPT during the data analysis phase. They asked ChatGPT \textit{``How to convert kelvin to celcius"}, with the follow-up \textit{``how many degrees celdius is 273 kelvin"}.
In the first response, ChatGPT provided the formula $T_C = T_K - 273.15$, and also included an example where $T_K = 300 \text{ K}$.
In the second response, ChatGPT said that with $T_K = 273 \text{ K}$, $T_C = -0.15 \text{ °C}$.
After the replies, the group went on to have a discussion without ChatGPT about whether they needed the temperature data or not for the experiment.
Here, the group still had no equation or conception of a model that included the temperature dependency for the speed of sound, making the interaction episode rather irrelevant.
The students' focus was locked on the aforementioned enacted object of learning, and they had yet to succeed with the first part of the analysis, which was to calculate the wavelength.
However, with no apparent understanding of how the speed of sound, or how the concept of wavelengths were connected to the experimental setup, Group 2 would have needed clearer guidance towards aspects relevant to the object of learning.
Instead, here the group's interaction with ChatGPT strengthened their focus on temperature, and they kept on haphazardly trying various calculations to finish the lab.

In episode two, Group 5 also struggled with the analysis and getting an estimate of the speed of sound using the collected data.
Towards the end of the lab session, they turned to ChatGPT and asked \textit{``what is speed of sound in 22 degrees celsius"}.
The reply told them that the approximate speed of sound at $22 \text{ °C}$ is $344 \text{ m/s}$, and the group was then happy to not continue with their own calculations but simply agreed that the answer given by ChatGPT was a good estimation of what they would have ended up with.
During the group's short discussion following this interaction, they said that the reply was reasonable since ChatGPT provides \textit{``a good average of previous research results"} (Eddie, Group 5), and the answer could thus be understood as ChatGPT had done the analysis for them.
Here we could identify a similar enacted object of learning, albeit more closely connected to simply being able to provide an acceptable value for the speed of sound in air.
Similar to Group 2, Group 5 showed no prior understanding of the speed of sound or the connection to the concept of wavelength.
Drawing upon the identified enacted object of learning, together with the fact that Group 5 saw tools like ChatGPT as being able to provide a synthesis of all available information, their way of interacting with ChatGPT was perfectly reasonable.
Again, here we see yet another episode where the students' structure of awareness, both regarding the object of learning and GenAI tools, is not affected by spontaneous interactions with ChatGPT.

Episode three consists of an interaction Group 4 had with ChatGPT towards the end of the lab session. There, they discussed how particles could be levitated using the experimental setup together with ChatGPT using the prompts \textit{``how can sound waves counteract gravity"}, \textit{``what do you mean by the gravitational force is balanced?"}, \textit{``simplify how sound waves counteract the gravitational force ?"}, and \textit{``is it particles that have bounced on each other and in that way creates a force upward that takes out the gravitational force"}.
In between the prompts, Group 4 tried to identify if the answers they got agreed with their prior knowledge and results from the experiment.
For Group 4, we identified a clear contrast to the previous two episodes, namely that the enacted object of learning is more closely connected to the full stated aim in the lab manual.
This is evident in part due to the fact that the group thought it important to not only calculate the speed of sound, but also to understand connections to the phenomenon of acoustic levitation.
Another contrast is seen in the way Group 4 interacted with ChatGPT.
Here, the group's discussions prior to and between the first two interactions indicate that they initially have discerned a connection between the concept of wavelength and the levitating styrofoam particles.
However, the students perceive that ChatGPT provides confounding information and introduces concepts and ideas they are currently unfamiliar with.
Additionally, their lack of adequate knowledge about the underlying physics makes it hard for them to gauge the correctness of the answers ChatGPT provides.
Thus, the interaction led to a shift of awareness regarding critical aspects of acoustic levitation with the equipment; from having to do with sound waves from the two speakers sometimes \textit{``strengthening"} each other (I.e., on their way to discovering standing waves) and towards one students original notion that the styrofoam levitates mainly due to \textit{``air particles"} physically pushing on it from below.

\section{Discussion}
\label{sec:discussion}

The findings in the current study are based on seven lab groups and the work and reflection of, in total, 19 students.
As such, they should not be interpreted as directly generalizable. However, this study provides important insights into the potential use, current and future, of GenAI tools in high school science laboratories.
Here, we discuss the findings in relation to the research questions, as well as broader implications of using GenAI tools in laboratory work.
The discussion is distilled into four parts: \textit{The role of prior knowledge and trustworthiness}, \textit{Utilizing ChatGPT as a lab partner or a source of information}, \textit{GenAI as a tool to develop problem-solving skills in laboratory settings}, and \textit{The ongoing problem with GenAI and hallucination}.

\subsection{The role of prior knowledge and trustworthiness}

Regarding RQ1: \textit{What do students perceive as useful interactions with GenAI tools like ChatGPT during a lab session}, our findings find that, in agreement with previous research \cite{Dahlkemper2023}, students require knowledge about the underlying physics in order for them to gauge whether answers provided by ChatGPT were trustworthy.
This is exemplified well by Group 6, one of two groups of third-year students taking the optional course Physics 3, where they could get meaningful assistance from ChatGPT during the analysis stage of the lab in identifying a formula to calculate the speed of sound that included temperature dependence.
If not for the group's prior knowledge of the physics of sound, they might not have been able to conclude that the initial equation provided by ChatGPT was faulty in that the form of the equation was correct, but the gas constant $R$ was given with the wrong units and would have had to be converted prior to using it to calculate the speed of sound.

The importance of prior knowledge about the specific area to successfully identify the validity of answers given by ChatGPT is further strengthened by examining an interaction during the lab by Group 4.
Although they were one of the groups with second-year students taking Physics 1 and thus having access to very little prior knowledge of the physics of sound, they were identified as performing well during the lab since they were able to identify what data they should collect and perform the analysis without issues.
However, they ended up struggling towards the end of the lab session when they were tasked to discuss how acoustic levitation works. Due to their limited to non-existent knowledge of wave mechanics and the physics of sound, they had few tools available to them to challenge what answers ChatGPT gave them.
As a result, despite Group 4 demonstrating an understanding of the need to pose not just one question but many to try to identify if the answers provided by ChatGPT made sense or not, they were ultimately led down a path toward a wrong explanation that it is mainly the lower speaker that supplies a force by pushing on the styrofoam particle and thus counteracting gravity.

Finally, in most other interactions with ChatGPT during the lab session, the groups seemed to take the answers provided by ChatGPT as simply being the truth.
This may be connected to how the students, at times, anthropomorphized ChatGPT during the interviews.
Group 3 stated that they might not fully trust the answers if ChatGPT did not provide any sources or citations, and Group 5 stated that ChatGPT can be trusted since it has knowledge about previous research results.
In general, during the group interviews before and after the lab, when asked about the capabilities of ChatGPT, students were quick to use words such as knowledge and talk less about it as some tool and more as a person with much knowledge.
During the interviews, it was mainly groups 6 and 7 that stood by the argumentation that what ChatGPT can answer and not reliably is dictated by what data was part of its training, which indicates that they are more cognizant that GenAI tools do not have knowledge and reasoning skills similar to humans.

\subsection{Utilizing ChatGPT as a lab partner or a source of information}

Concerning RQ2 and RQ3, they focus more on how students interacted with ChatGPT during the lab to overcome specific hurdles.
Previous studies provide little information about its capabilities to function as a lab partner.
What it does show, however, is that depending on the type of questions and problems students face, it might be able to provide reliable answers, as it can be good at solving subject-specific problems \cite{Kortemeyer2023, lopez2024}.
Further, it has also been identified that GenAI tools, like ChatGPT, can reproduce language and provide answers relevant to tackling problems in laboratory settings \cite{arujo2024}.
Our findings indicate similar conclusions. In RQ2, \textit{How do high school students actively engage with ChatGPT to solve problems during a lab investigating the speed of sound using acoustic levitation}, Fig. \ref{fig:gantt} shows that students tend to make use of ChatGPT after having done all or most of the data collection.
This indicates that despite having little to no knowledge about the phenomenon under study, the groups had little problem collecting data using the experimental setup.

During the group interviews, students stressed that a major use case of ChatGPT is to get answers to conceptual questions or to clarify problems or instructions, and that it made little or no sense in asking about lab equipment.
This sentiment is in agreement with how the groups did make use of ChatGPT during the lab session: during the analysis and to get help with the discussion questions about acoustic levitation.
Here, the primary beneficial use case for the groups was related to getting help with identifying what equations to make use of, as well as getting an introduction to acoustic levitation.
A noteworthy special case was Group 5, who \textit{``outsourced"} the analysis to ChatGPT after failing to understand what to do with the data by simply asking what the sound speed was at the room's current temperature.
During the analysis, it is important to stress that the lab instructions were intentionally vague in that no equations were provided.
Instead, the assumption was that students would utilize a combination of ChatGPT and their formula book to identify how to make use of the data to get the speed of sound using the equation $v = f \lambda$ and compare that to the speed of sound they get when using an equation that included temperature dependence.
Additionally, if the groups had time, they were tasked to discuss questions related to acoustic levitation, including the underlying principles of acoustic levitation.
As such, it was expected that all groups, regardless of their natural intention to use or not use ChatGPT during the study, were incentivized to at least try to use it during the analysis and discussion part.

\subsection{GenAI as a tool to develop problem-solving skills in laboratory settings}

Overall, the findings presented in the current study demonstrate fruitful and problematic interactions with ChatGPT in terms of aiding students in solving problems or helping them understand the phenomenon of acoustic levitation.
Groups 4 and 6 were both able to use ChatGPT to aid them during the analysis in that Group 4 successfully utilized it to aid them in reasoning about whether the wavelength they got from the data collection was reasonable.
Further, both groups showed proficiency in discussing how to identify if the equations given to them from ChatGPT were reliable by drawing upon their prior knowledge of dimensional analysis.
Contrary to these, groups 1, 2, 3, and 5 all demonstrated that lacking knowledge about the relevant physics and how GenAI tools like ChatGPT \textit{``reasons"} or has \textit{``knowledge"}, having access to such tools in a classroom setting, including laboratories, might end up disadvantageous for the students learning process.

In this contrast between groups 4 and 6, and groups 1, 2, 3, and 5, we can see a situation where ChatGPT can serve as a tool to help reduce the importance of prior content knowledge during lab sessions, which teachers can utilize to help their students develop certain capabilities.
Namely, that groups 4 and 6, despite significant differences regarding relevant knowledge of the underlying physics, both were able to make use of ChatGPT to assist them in moving forward with the data analysis.
What instead separated these groups from 1, 2, 3, and 5, was their assumptions and knowledge about how GenAI tools function and \textit{``reasons."}
The kind of interactions groups 4 and 6 demonstrated during the analysis provided them with new tools on how to approach problem-solving in a laboratory context.
Drawing upon variation theory, we can identify an alternative approach to the lab session that could allow the struggling groups to succeed more similarly to groups 4 and 6.
If an object of learning is chosen related to utilizing GenAI tools in problem-solving during laboratory work, a potential critical aspect may be related to discerning how current GenAI tools' reasoning differs from that of humans.
Here, the argument for the critical aspect is connected to the identified stark difference between groups 4, 6, and 7 and the other groups when it came to assumptions about how GenAI tools work.
Such a focus would also open up for more active scaffolding regarding steering the students' focus towards critical aspects \cite{Kullberg2024}.
This would include rather heavily aiding students with content-specific conceptual understanding, leaving them to instead focus on problem-solving and GenAI-related aspects.
Finally, this shift in focus towards more general skills has been identified in previous work in PER using variation theory, as exemplified by \cite{KildeLfgren2023} on how students develop skills related to developing internal idealized models and use them to develop conceptual understanding in the physics laboratory (referred to as being able to modelize). It is also a focus that has widespread acceptance in the current PER literature as a means towards getting a more expert-like view of laboratory work in physics (e.g., \cite{Etkina2010, Etkina2010b, Wilcox2017, Holmes2018}).

Continuing, we can connect the suggested change in focus regarding the object of learning to RQ3, \textit{How can interacting with ChatGPT during the lab aid students in solving problems and with conceptual understanding}.
With the aim of having students develop new skills using GenAI when solving problems during laboratory work, a crucial part of their scaffolding during the lab would revolve around probing the students on their conceptual understanding of the phenomenon studied during the lab by actively interacting with the groups during the lab.
We base this suggestion on that our findings show instances of ChatGPT providing information to the students about the underlying physics, which is wrong or misleading.
An example of such an instance is seen in the interaction Carl in Group 3 has with ChatGPT.
There, the analogy of keeping a ball in the air by blowing on it from below reinforced the notion that the main contributor to the phenomenon seen during the lab was due to the bottom speaker pushing air particles on the styrofoam particle.
If a teacher were present and identified Carl's acceptance of this analogy, they could intervene and provide an explanation to correct the problems with the analogy.
Thus, Carl and the rest of Group 3 would be back towards the intended object of learning instead of having their awareness shifted towards aspects related to a more concept-focused object of learning.

\subsection{The ongoing problem with GenAI and hallucinations}

Becoming better at addressing the issue of getting problematic and made-up answers from GenAI tools in the classroom relates to the growing need for AI literacy in education, for students and teachers alike \cite{Ding2023, Lo2024, Polverini2024}.
As is evident from our findings, and in agreement with previous work \cite{Xia2023}, students with a combination of less knowledge of the subject (here, physics) and not being familiar with GenAI tools or have little to no knowledge about how they function struggle with making use of these new tools.
Proposed solutions from the literature involve training students and teachers in prompt engineering \cite{Ding2023, Polverini2024, Wang2024-IEEE}.
This proposal is further supported by our findings, where groups 4 and 6 utilized their understanding and assumptions about how ChatGPT functions to help them distinguish between reliable and unreliable information in the answers provided by the chatbot.
From their interviews and observations of their lab sessions, it became evident that their assumptions led them to ascribe a less human-like awareness in how ChatGPT generated responses.

Developing skills related to prompt engineering can be a fruitful way to give students and teachers a greater understanding of not only how to extract more useful information from GenAI tools, but also to better comprehend changes in limitations over time between and within models \cite{Ding2023, Polverini2024}.
For educators, an important step in learning how the behavior of GenAI tools varies over time is for researchers to continuously monitor how various models perform, with and without proper prompt engineering and further fine-tuning.
To that end, the current study has explored how largely untrained students can and do make use of a free version of ChatGPT, GPT-3.5 at the time of data collection, when solving problems during a physics lab.
The findings and discussions presented here can benefit educators and researchers in gaining a deeper understanding of the naïve and partially spontaneous interaction between students and GenAI, particularly in identifying fruitful ways of interaction and problematic instances of overreliance on ChatGPT as a substitute for or absolute source of information.

The use of GPT-3.5, rather than, for example, GPT-4o, has profound implications in terms of the type of misinformation presented by ChatGPT, as well as how it responds to questions in this study.
Additionally, it is reasonable to assume that the way ChatGPT behaves has an effect on how participants in this study rely on the answers and find ChatGPT helpful.
However, subsequent releases of GenAI tools still struggle with the problem of hallucinations, albeit less frequently, and it is a problem that may be inherent to the way large language models work \cite{Xu2025}.
This implies that although the specific problems ChatGPT exhibits in this study may not be present if replicating the study today, we would expect similar conclusions, but perhaps that more groups may have been able to successfully use ChatGPT, related to when GenAI tools can be useful and why when considering what is to be learned.

\section{Conclusion and outlook}
\label{sec:conclusion}

GenAI has made a clear impact on the public and educational discourse about the future of work and education.
It, and AI more generally, are looking to be part of people's daily lives to an even larger extent in the future, both at an individual and professional level.
Therefore, it is essential that, as far as possible, both teachers and students are instructed in how to interact with GenAI tools to extract the greatest benefits from them.
This has happened with other technologies in the past, including calculators, digital circuits, computers, and the internet.
However, GenAI tools are different in that they respond to the user in fundamentally new ways, thus requiring new kinds of training related to formulating questions and gauging the reliability of results and answers from such tools.

In the current study, we address the research gap related to the need of exploring how GenAI tools, specifically the free version ChatGPT at the time of the data collection, could be used in the educational physics laboratory.
Our findings are in agreement with previous research regarding the importance of educating students about the capabilities and limitations of AI tools, as well as gaining some level of understanding about how they are developed and \textit{``think."}
Additionally, in our analysis, we have explored a novel way of conceptualizing the enacted object of learning from a student perspective, and its connection to the structure of awareness.
This allowed us to gain further insight into how GenAI tools can be utilized in laboratory settings to develop students' problem-solving skills.

In STEM education specifically, there is a need to further explore and develop educational strategies and material that can help students and teachers better understand how to best utilize and think about the role of GenAI in education, not the least in laboratory settings.
As these tools continue to be developed, their capabilities to simulate a trustworthy chain of reasoning will improve, furthering the need to carefully consider what implications interactions with such tools may have when it comes to learning concepts in disciplinary-accepted ways.
Here, the role and knowledge of the teacher about AI is of utmost importance. Finally, we have identified that there is a potential to make use of GenAI tools like ChatGPT in the physics laboratory as a lab partner, but that the usefulness looks to be dependent on students knowledge about the phenomenon under study as well as their knowledge about how GenAI tools function.
Furthermore, the specific ways in which such a tool is best employed in the physics laboratory require more research and will also be dictated by the future development of such tools.

\begin{acknowledgments}
We would like to thank the students who participated in the study.
We also thank our dear friend and colleague Dr. Ricardo Méndez-Fragoso for providing valuable insights in discussing and reviewing the manuscript. 

\subsection*{Author contributions}

\textbf{Sebastian}: Conceptualization, Data curation (lead), Formal analysis, Investigation (lead), Methodology, Project administration, Resources (lab design; supporting), Validation (lead), Visualization (lead), Writing - original draft, Writing - review \& editing (lead). \\
\textbf{Andreas}: Investigation (supporting), Resources (lab design; lead), Validation (supporting), Visualization (supporting), Writing - original draft (supporting), Writing - review \& editing (supporting). \\
\textbf{Jonas}: Data curation (supporting), Investigation (supporting), Supervision, Validation (supporting), Writing - review \& editing (supporting).
\end{acknowledgments}

\newpage
\appendix
\label{appx:manual}
\section{Lab Manual}
\section*{Speed of sound and wavelength}

\subsection*{Aim}
To explore acoustic levitation using \textbf{LeviLab} and determine the speed of sound in air.

\subsection*{Equipment}
\begin{itemize}
    \item LeviLab (two ultrasonic speakers mounted on a caliper, a function generator, an amplifier, and a thermometer)
    \item Lab stand equipment 
    \item Voltage supply (DC, 20 V)
    \item Polystyrene particles
    \item ChatGPT
\end{itemize}

\subsection*{Experiment}
\begin{enumerate}
    \item Information, assembly, and operation of LeviLab
    \begin{itemize}
        \item Place LeviLab on a flat surface.
        \item Stabilize LeviLab using the lab stand equipment.
        \item Vary the loudspeaker distance using the caliper.
        \item The emitted sound frequency is 40 kHz.

    \end{itemize}
    \item Calculation of the speed of sound in air
    \begin{itemize}
        \item Verify that the power supply is on.
        \item Place a polystyrene particle on the lower speaker.
        \item Gradually lower the upper loudspeaker and identify points where the particle appears to `jump up' and start levitating. Does increasing the distance between the speakers produce the same effect?
        \item Place the loudspeakers at a distance from each other so that as the loudspeaker distance approaches zero, the particle has jumped 7-10 times. Here is a closer description of a measurement:
        \begin{itemize}
            \item Carefully lower the top loudspeaker. Stop exactly at the position where the polystyrene particle jumps up and starts to levitate. 
            \item Use the caliper to record the distance between the loudspeakers.
            \item Record the temperature of the ambient air using the thermometer.

        \end{itemize}

    \end{itemize}
    \item Analysis
    \begin{itemize}
        \item Calculate the wavelength
        \begin{itemize}
            \item Label your measurements such that 1 represents the shortest speaker spacing. This is called an ordinal number.
            \item Conduct a linear regression analysis with loudspeaker spacing as the dependent variable and the ordinal number as the independent variable. The slope of the regression line corresponds to the wavelength.

        \end{itemize}
        \item Calculate the speed of sound using a formula involving wavelength and frequency.
        \item Compare your result with the speed of sound as determined by the temperature reading. This is achieved by applying a linear model that expresses the relationship between the speed of sound and the temperature in degrees Celsius.

    \end{itemize}
\end{enumerate}
\subsection*{Discuss in the group}
\begin{enumerate}
    \item What are the fundamental principles of acoustic levitation that make it possible for sound waves to balance the gravitational force on an object? Construct a descriptive figure.
    \item Discuss possible sources of error and how to minimize them to calculate more reliable speed of sound values.
    \item Explain why it is important to measure the ambient temperature when calculating the speed of sound.
    \item Discuss potential applications of acoustic levitation in situations outside school.
    \item Explore and describe other applications where standing waves are used.

\end{enumerate}

\bibliography{refs.bib}

\end{document}